\documentclass[preprint,showpacs,preprintnumbers,amsmath,amssymb]{revtex4}

\usepackage{graphicx}
\usepackage{dcolumn}
\usepackage{bm}

\begin{document}


\title{A novel method to construct stationary solutions of the Vlasov-Maxwell system : the relativistic case}

\author{Akihiro Suzuki}
\email{suzuki@resceu.s.u-tokyo.ac.jp}
\affiliation{Research center for the Early Universe, Graduate School of Science, University of Tokyo, Bunkyo-ku, Tokyo 113-0033, Japan}
\altaffiliation[Also at ]{Department of Astronomy, Graduate School of Science, University of Tokyo, Bunkyo-ku, Tokyo 113-0033, Japan}

\date{\today}

\begin{abstract}
A method to derive stationary solutions of the relativistic Vlasov-Maxwell system is explored. 
In the non-relativistic case, a method using the Hermite polynomial series to describe the deviation from the Maxwell-Boltzmann distribution is found to be successful in deriving a few stationary solutions including two dimensional one. 
Instead of the Hermite polynomial series, two special orthogonal polynomial series, which are appropriate to expand the deviation from the Maxwell-J\"uttner distribution, are introduced in this paper. 
By applying this method, a new two-dimensional equilibrium is derived, which may provide an initial setup for investigations of three-dimensional relativistic collisionless reconnection of magnetic fields. 
\end{abstract}

\pacs{Valid PACS appear here}
\maketitle

\section{\label{sec:level1}INTRODUCTION}
Relativistic plasmas play important roles in high energy phenomena. 
In nature, gamma-ray burst afterglow is considered to be a phenomenon resulting from interaction of ultrarelativistic plasmas with a Lorentz factor $\gamma\simeq 10^2-10^3$ \cite{SP90,Paczynski}. 
Detailed observations of high energy astrophysical phenomena, such as supernova remnants, gamma-ray bursts, and active galactic nuclei, reveal that non-thermal components in their spectrum comprise of radiation from relativistic particles. 
In recent laser experiments, interaction of laser and plasmas is realized in highly relativistic regime. 
These phenomena often involve sufficiently rarefied plasma that the effect of collisions is negligible. 
In order to give plausible explanations for such phenomena, many authors have devoted their energies to investigations into properties of relativistic plasmas, i.e., instabilities, radiation, and stationary equilibria. 

Here I propose a new method to obtain stationary equilibrium configurations of collisionless plasmas described by the relativistic Vlasov-Maxwell system. 
Such solutions are of great interest because they give self-consistent configurations of relativistic plasmas, which are used as an initial setup for studies on the relativistic collisionless reconnection of magnetic fields \cite{parker,sweet,petschek}. 
In some recent numerical simulations of the relativistic collisionless magnetic reconnection \cite{ZH01,J04}, relativistic extensions of the harris sheet equilibrium \cite{harris} are used as the initial setup. 
While stationary solutions of the non-relativistic Vlasov-Maxwell system are well explored \cite{harris,bennett,BGK57,mahajana,mahajanb,AP99,BBSS01,mottez,CMPC05,MP07}, exploration of the relativistic Vlasov-Maxwell equilibria is not sufficient. 
Although Ref. \cite{Braasch} discussed the existence of such solutions and derived a special class of them, studies on the derivation of the concrete expressions of the relativistic Vlasov-Maxwell equilibria are still rare. 

In the previous paper \cite{SS08}, a novel method to construct stationary solutions of the non-relativistic Vlasov-Maxwell system is proposed. 
In the method, the key concept is to describe the deviation of the distribution function from the Maxwell-Boltzmann by orthogonal polynomial series. 
The Hermite polynomial series is chosen for the purpose, because the weight function is Gaussian, which is equivalent to the Maxwell-Boltzmann distribution. 
Ref. \cite{SS08} reveals that this choice is appropriate to recover some well-known equilibria (the Harris sheet and the Bennet pinch) and derive a new equilibrium. 
In this paper, I will extend the previous method to deal with relativistic plasmas. 
Instead of the Maxwell-Boltzmann distribution, I use  a relativistic extension of the Maxwell-Boltzmann distribution, i.e., the Maxwell-J\"uttner distribution \cite{J11} as the weight function of these polynomial series. 
To the best of my knowledge, there is no appropriate orthogonal polynomial series that describes the deviation from the Maxwell-J\"uttner distribution as the Hermite polynomial does for non-relativistic plasmas. 
So I introduce two orthogonal polynomials in this paper. 
Applying the method, I can derive a new two-dimensional equilibrium, which is a relativistic extension of the equilibrium proposed in the previous paper. 
This equilibrium may provide an initial setup for numerical simulations of the relativistic collisionless reconnection. 

This paper is organized as follows. 
In Sec.\ref{form}, I describe the procedure of the method. 
In Sec.\ref{ex}, I apply the method to derive a stationary solution that I can treat it analytically. 
I conclude this paper in Sec.\ref{conc}. 
Appendix gives some algebraic relations used in this paper.

\section{\label{form}FORMULATION}
The relativistic Vlasov equation describes the kinetic evolution of the distribution function $f_j(t,\mathbf{x},\mathbf{p})$ of particles $j$ ($=i$ for ions and $e$ for electrons) in the phase space $(x,y,z,p_x,p_y,p_z)$ and the Maxwell equations describe the evolution of the electromagnetic fields. 
Here the cartesian coordinates in the real space are $(x,y,z)$ and the corresponding coordinates in the momentum space are $(p_x,p_y,p_z)$. 
This system describes the exact behavior of relativistic collisionless plasmas. 
In this section, I derive stationary configuration of plasmas uniformly extending in the $z$ direction governed by the relativistic Vlasov-Maxwell system under the assumptions described in the following subsection. 
\subsection{Equations}
Because physical variables do not depend on $t$ or $z$, the relativistic Vlasov equation is expressed as
\begin{equation}
\begin{aligned}
&\frac{p_x}{\sqrt{m_j^2c^2+|\mathbf{ p}|^2}}\frac{\partial f_j}{\partial x}+
\frac{p_y}{\sqrt{m_j^2c^2+|\mathbf{ p}|^2}}\frac{\partial f_j}{\partial y}+
\frac{q_j}{c}\left(E_x-\frac{p_z}{\sqrt{m_j^2c^2+|\mathbf{p}|^2}}B_y\right)
\frac{\partial f_j}{\partial p_x}\\
&\ \ \ +\frac{q_j}{c}\left(E_y+\frac{p_z}{\sqrt{m_j^2c^2+|\mathbf{p}|^2}}B_x\right)
\frac{\partial f_j}{\partial p_y}
+\frac{q_j}{c}\left(\frac{p_x}{\sqrt{m_j^2c^2+|\mathbf{p}|^2}}B_y-\frac{p_y}{\sqrt{m_j^2c^2+|\mathbf{p}|^2}}B_x\right)\frac{\partial f_j}{\partial p_z}
=0, 
\label{vlasov}
\end{aligned}
\end{equation}
where $f_j$ is the distribution function for particles $j$ with the charge $q_j$ and the mass $m_j$, $c$ is the speed of light and $|\mathbf{p}|=\sqrt{p_x^2+p_y^2+p_z^2}$ is the norm of the momentum. 
$E_x,E_y$ and $B_x,B_y$ represent the $x,y$-components of electric and magnetic fields, which are functions of $x$ and $y$.
These electromagnetic fields are written as
\begin{subequations}
  \begin{eqnarray}
    E_x=-\frac{\partial \phi}{\partial x}&,&E_y=-\frac{\partial \phi}{\partial y},\\
    B_x=\frac{\partial A_z}{\partial y}&,&B_y=-\frac{\partial A_z}{\partial x},
  \end{eqnarray}
  \label{pot}
\end{subequations}
by introducing the scalar potential $\phi(x,y)$ and the $z$-component of the vector potential $A_z(x,y)$. 
The other components are assumed to vanish. 
These potentials satisfy the Poisson equations;
\begin{subequations}
  \begin{eqnarray}
    \frac{\partial^2\phi}{\partial x^2}+\frac{\partial^2\phi}{\partial y^2}&=& 
    -4\pi\rho,\\
    \frac{\partial^2A_z}{\partial x^2}+\frac{\partial^2A_z}{\partial y^2}&=& 
    -4\pi j_z,
  \end{eqnarray}
  \label{poisson}
\end{subequations}
where $\rho(x,y)$ and $j_z(x,y)$ are the charge density and the $z$-component of the electric current density, respectively. These are expressed in terms of $f_j(x,y,p_x,p_y,p_z)$ as
\begin{subequations}
  \begin{eqnarray}
    \rho&=&\sum_jq_j\int^\infty_{-\infty}dp_x\int^\infty_{-\infty}dp_y\int^\infty_{-\infty}dp_zf_j,\\
    j_z&=&\sum_jq_j\int^\infty_{-\infty}dp_x\int^\infty_{-\infty}dp_y\int^\infty_{-\infty}dp_z\frac{p_z}{\sqrt{m_j^2c^2+|\mathbf{p}|^2}}f_j,
  \end{eqnarray}
  \label{current}
\end{subequations}
which close the system. 
Solutions of Equations (\ref{vlasov})-(\ref{current}) give self-consistent configurations of collisionless plasmas.

\subsection{Derivation of stationary solutions}
In the following, I describe the procedure to derive solutions of Equations (\ref{vlasov})-(\ref{current}). 
The non-relativistic counterpart derived in the previous paper \cite{SS08} was based on the assumption that the deviation from the Maxwell-Boltzmann distribution can be expanded by the Hermite polynomial series. 
For a relativistic plasma, the distribution function in the thermodynamical equilibrium is given by the Maxwell-J\"uttner distribution \cite{J11}, a relativistic extension of the classical Maxwell-Boltzmann distribution;
\begin{equation}
f^\mathrm{MJ}_j=\frac{n_j}{4\pi m_j^2ck_\mathrm{B}T_jK_2(\zeta_j)}
\exp\left[-\zeta_j\sqrt{1+\hat{p}_x^2+\hat{p}_y^2+\hat{p}_z^2}\right],
\end{equation}
where $k_\mathrm{B}$ is the Boltzmann constant. 
$T_j$ and $n_j$ represents the temperature and the density of the particles $j$. 
I have assumed they are constant. 
$K_2(\zeta_j)$ is the modified Bessel function of the second kind. 
Some dimensionless variables have been introduced as $\zeta_j=m_jc^2/(k_\mathrm{B}T_j)$, $\hat{p}_x=p_x/(m_jc)$, $\hat{p}_y=p_y/(m_jc)$, and $\hat{p}_z=p_z/(m_jc)$. 

In seeking solutions of Equations (\ref{vlasov})-(\ref{current}), I assume that the distribution function takes the following form;
\begin{equation}
  f_j=\left\{\sum_{n=0}^\infty \left[g^\mathrm{even}_{j,n}(A_z)S^\mathrm{even}_n(\hat{p}_z)+ g^\mathrm{odd}_{j,n}(A_z)S^\mathrm{odd}_n(\hat{p}_z)\right]\right\}
\exp\left(-\frac{q_j\phi}{k_\mathrm{B}T_j}\right)f^\mathrm{MJ}_j,
  \label{fj}
\end{equation}
where $S_n^\mathrm{even}$ and $S_n^\mathrm{odd}$ are orthogonal polynomial series defined in Appendix \ref{Orth}. 
Their coefficients $g_{j,n}^\mathrm{even}$ and $g_{j.n}^\mathrm{odd}$ are assumed to be functions of $A_z$. 
The distribution function in this form describes the deviation from the Maxwell-J\"uttner distribution in terms of the orthogonal polynomial expansion. 
The spatial dependence of the distribution function is described through those of potentials $\phi$ and $A_z$. 
For large $\zeta_j$, these polynomials can be reduced to the Hermite polynomial series as
\begin{equation}
S^\mathrm{even}_n=\frac{H_{2n}(\sqrt{2\zeta}\hat{p}_z)}{\zeta^n},\ \ \ 
S^\mathrm{odd}_n=\frac{H_{2n+1}(\sqrt{2\zeta}\hat{p}_z)}{\zeta^{n+1/2}}.
\end{equation}
Thus this method becomes equivalent to the previous method \cite{SS08} in the non-relativistic limit. 

At first, substituting the expressions for fields (\ref{pot}) and the distribution function (\ref{fj}) into Equation (\ref{vlasov}), we obtain
\begin{equation}
\left\{\sum_{n=0}^\infty \left[
\frac{dg_{j,n}^\mathrm{even}}{dA_z}S^\mathrm{even}_n(\hat{p}_z)+
\frac{dg_{j,n}^\mathrm{odd}}{dA_z}S^\mathrm{odd}_n(\hat{p}_z)-
\frac{q_jg_{j,n}^\mathrm{even}}{m_jc^2}\frac{dS_n^\mathrm{even}}{d\hat{p}_z}-
\frac{q_jg_{j,n}^\mathrm{odd}}{m_jc^2}\frac{dS_n^\mathrm{odd}}{d\hat{p}_z}
\right]\right\}f^\mathrm{MJ}_j=0
\label{integrand}
\end{equation}
Then, to obtain equations to determine the coefficients $g^\mathrm{even}_{j,n}$ and $g^\mathrm{odd}_{j,n}$, both sides of this equation is multiplied by $S_m^\mathrm{even}(\hat{p}_z)$ and integrated with respect to $\hat{p}_x$, $\hat{p}_y$, and $\hat{p}_z$. 
The orthogonality relation of $S^\mathrm{even}_{n}$ (\ref{orth_even}), the relation (\ref{vanisha}), and the expression (\ref{dSodd}) yield the following ordinary differential equations,
\begin{equation}
\frac{dg^\mathrm{even}_{j,m}}{dA_z}-
\frac{q_j}{m_jc^2}\sum_{k=m}^\infty B_{km}^\mathrm{odd}g^\mathrm{odd}_{j,k}=0.
\label{releven}
\end{equation}
On the other hand, multiplying both sides of Equation (\ref{integrand}) by $S_m^\mathrm{odd}(\hat{p}_z)/\sqrt{1+\hat{p}_x^2+\hat{p}_y^2+\hat{p}_z^2}$ and integrating with respect to $\hat{p}_x$, $\hat{p}_y$, and $\hat{p}_z$, 
other equations for coefficients
\begin{equation}
  \frac{dg^\mathrm{odd}_{j,m}}{dA_z}-
  \frac{q_j}{m_jc^2}\sum_{k=m+1}^\infty B_{km}^\mathrm{even}g^\mathrm{even}_{j,k}=0,
  \label{relodd}
\end{equation}
are obtained. 
Here the relations (\ref{orth_odd}) and (\ref{vanishb}) have been used. 
Equations (\ref{releven}) and (\ref{relodd}) give the relation between $g^\mathrm{even}_{j,n}$ and $g^\mathrm{odd}_{j,n}$. 
However, it is not practical to directly solve these infinite number of equations. 
Some restrictions are needed in order to solve them. 
For example, the condition $g^\mathrm{even}_{j,n}=g^\mathrm{odd}_{j,n}=0 (n>N)$ or $g^\mathrm{odd}_{j,n}=g^\mathrm{even}_{j,n+1}=0 (n>N)$, reduces Equations (\ref{releven}) and (\ref{relodd}) to a finite number of ordinary differential equations and enable us to obtain expressions of $g^\mathrm{even}_{j,n}$ and $g^\mathrm{odd}_{j,n}$ as functions of $A_z$. 

To derive the source terms in the Maxwell equations, substituting Equation (\ref{fj}) into Equations (\ref{current}) and using the orthogonality relations (\ref{orth_even}), (\ref{orth_odd}), and (\ref{vanish}) again, these terms are written in terms of the potentials as
\begin{subequations}
  \begin{eqnarray}
    \rho&=&n_iq_ig_{i,0}^\mathrm{even}
    \exp\left(-\frac{q_i\phi}{k_\mathrm{B}T_i}\right)
    +n_eq_eg_{e,0}^\mathrm{even}
    \exp\left(-\frac{q_e\phi}{k_\mathrm{B}T_e}\right),\\
    j_z&=&\frac{n_iq_i}{\zeta_i}g^\mathrm{odd}_{i,0}
    \exp\left(-\frac{q_i\phi}{k_\mathrm{B}T_i}\right)+
    \frac{n_eq_e}{\zeta_e}g^\mathrm{odd}_{i,0}
    \exp\left(-\frac{q_e\phi}{k_\mathrm{B}T_e}\right),
  \end{eqnarray}
\label{rhoj}
\end{subequations}
where the expressions (\ref{Iodd}) and (\ref{Ieven}) have been used for evaluation of integrals. 
Substitution of these expressions into Equations (\ref{poisson}) leads to 
\begin{subequations}
  \begin{eqnarray}
    \frac{\partial^2\phi}{\partial x^2}+\frac{\partial^2\phi}{\partial y^2}&=& 
    -4\pi\left[n_iq_ig_{i,0}^\mathrm{even}
    \exp\left(-\frac{q_i\phi}{k_\mathrm{B}T_i}\right)
    +n_eq_eg_{e,0}^\mathrm{even}
    \exp\left(-\frac{q_e\phi}{k_\mathrm{B}T_e}\right)\right],
    \label{poisson2a}\\
    \frac{\partial^2A_z}{\partial x^2}+\frac{\partial^2A_z}{\partial y^2}&=& 
    -4\pi\left[\frac{n_iq_i}{\zeta_i}g^\mathrm{odd}_{i,0}
    \exp\left(-\frac{q_i\phi}{k_\mathrm{B}T_i}\right)+
    \frac{n_eq_e}{\zeta_e}g^\mathrm{odd}_{e,0}
    \exp\left(-\frac{q_e\phi}{k_\mathrm{B}T_e}\right)\right].
    \label{poisson2b}
  \end{eqnarray}
  \label{poisson2}
\end{subequations}
Since $g^\mathrm{even}_{j,0}$ and $g^\mathrm{odd}_{j,0}$ are functions of $A_z$, these equations determine the potentials $\phi$ and $A_z$ as functions of $x$ and $y$ under certain boundary conditions.

As well as the previous method \cite{SS08}, there is a limit in this method. 
For a stationary equilibrium, distribution functions can take various forms as long as the pressure balance is achieved. 
However, Equations (\ref{poisson2}) look as if there is a one-to-one correspondence between the field configuration and the distributions $g_{j,0}^\mathrm{even}$ and $g_{j,0}^\mathrm{odd}$. 
This disagreement arises from the assumption that the deviation of the distribution function from the Maxwell-J\"uttner distribution can be expanded by the polynomial series $S^\mathrm{even}_n$ and $S^\mathrm{odd}_n$. 
Therefore, the stationary solutions derived above cover a part of many possible equilibria. 

\section{\label{ex}Application}
In this section, I consider an application of the stationary solutions derived above to a plasma in a charge neutrality comprising of electrons and ions with the same charge but the opposite sign ($q_i=-q_e=e$ and $n_i=n_e=n_0$) in which the electric field strength is sufficiently small ($\phi=0$). 
Then, Equation (\ref{poisson2a}) implies that ions and electrons have the same spatial distribution;
\begin{equation}
g^\mathrm{even}_{i,0}=g^\mathrm{even}_{e,0}.
\label{neutral}
\end{equation}

As mentioned in the previous section, to solve Equations (\ref{releven}) and (\ref{relodd}), I truncate both of the series as $g^\mathrm{even}_{j,n}=0(n\ge 2)$ and $g^\mathrm{odd}_{j,n}=0(n\ge 1)$. 
Then, Equations (\ref{releven}) and (\ref{relodd}) reduce to the following three ordinary differential equations;
\begin{equation}
\begin{aligned}
\frac{dg^\mathrm{even}_{j,0}}{dA_z}=&\frac{q_j}{m_jc^2}g^\mathrm{odd}_{j,0},\\
\frac{dg^\mathrm{odd}_{j,0}}{dA_z}=&\frac{2q_j}{m_jc^2}g^\mathrm{even}_{j,1},\\
\frac{dg^\mathrm{even}_{j,1}}{dA_z}=&0,
\end{aligned}
\end{equation}
which have solutions in the form of
\begin{equation}
\begin{aligned}
g^\mathrm{even}_{j,0}=&g^\mathrm{even}_{j,1}(0)\left(\frac{q_jA_z}{m_jc^2}\right)^2
+g^\mathrm{odd}_{j,0}(0)\frac{q_jA_z}{m_jc^2}+g^\mathrm{even}_{j,0}(0),\\
g^\mathrm{odd}_{j,0}=&2g^\mathrm{even}_{j,1}(0)\frac{q_jA_z}{m_jc^2}
+g^\mathrm{odd}_{j,0}(0),\\
g^\mathrm{even}_{j,1}=&g^\mathrm{even}_{j,1}(0),
\label{solution}
\end{aligned}
\end{equation}
where $g^\mathrm{even}_{j,0}(0)$, $g^\mathrm{odd}_{j,0}(0)$, and $g^\mathrm{even}_{j,1}(0)$ are constants of integration, which can take various values as long as the distribution function $f_j$ is positive at arbitrary points in the phase space and the condition (\ref{neutral}) is satisfied. 
For example, I assume that they take the following values;
\begin{equation}
\begin{aligned}
&g^\mathrm{even}_{i,0}(0)=C,\ \ \ g^\mathrm{odd}_{i,0}(0)=0,\ \ \ 
g^\mathrm{even}_{i,1}(0)=1,\\
&g^\mathrm{even}_{e,0}(0)=C,\ \ \ g^\mathrm{odd}_{e,0}(0)=0,\ \ \ 
g^\mathrm{even}_{e,1}(0)=\frac{m_e^2}{m_i^2},
\label{conditions}
\end{aligned}
\end{equation} 
where $C$ is a dimensionless constant. 
Some algebraic manipulations results in the distribution functions in the form of
\begin{subequations}
\begin{eqnarray}
f_i&=&\frac{n_0}{2\pi ck_\mathrm{B}T_iK_2(\zeta_i)}
\left[\frac{1}{m_i^2c^2}\left(p_z+\frac{eA_z}{c}\right)^2+
C-\frac{K_3(\zeta_i)}{\zeta_i K_2(\zeta_i)}\right]\nonumber\\
&&\hspace{10em}\times
\exp\left[-\frac{c}{k_\mathrm{B}T_i}\sqrt{m_i^2c^2+p_x^2+p_y^2+p_z^2}\right],\\
f_e&=&\frac{n_0}{2\pi ck_\mathrm{B}T_eK_2(\zeta_e)}
\left[\frac{1}{m_i^2c^2}\left(p_z+\frac{eA_z}{c}\right)^2+
C-\frac{m_e^2K_3(\zeta_e)}{m_i^2\zeta_eK_2(\zeta_e)}\right]\nonumber\\
&&\hspace{10em}\times
\exp\left[-\frac{c}{k_\mathrm{B}T_e}\sqrt{m_e^2c^2+p_x^2+p_y^2+p_z^2}\right].
\end{eqnarray}
\label{distribution}
\end{subequations}
These exactly satisfy the relativistic Vlasov equation (\ref{vlasov}). 
In order that the distribution function of ions $f_i$ takes positive values, the conditions
\begin{equation}
C\geq\frac{K_3(\zeta_i)}{\zeta_i K_2(\zeta_i)},\ \ \ 
C\geq\frac{m_e^2K_3(\zeta_e)}{m_i^2\zeta_eK_2(\zeta_e)}
\end{equation}
are required. 

Next, the equations governing the field configuration is deduced by 
substituting the solutions (\ref{solution}) and (\ref{conditions}) into Equation (\ref{poisson2b})
\begin{equation}
\frac{\partial^2A_z}{\partial x^2}+\frac{\partial^2A_z}{\partial y^2}=
-\frac{8\pi e^2}{c^2}\left(\frac{n_im_i}{\zeta_i}+\frac{n_em_e}{\zeta_e}\right)
A_z.
\end{equation}
This equation has a solution in the form of 
\begin{equation}
A_z=-B_0L\left[\cos(x/L)+\cos(y/L)\right],
\label{vector}
\end{equation}
where $L$ is a constant satisfying the following relation;
\begin{equation}
L=\frac{c}{2e}\sqrt{\frac{\zeta_i\zeta_e}
{2\pi(n_im_i\zeta_e+n_em_e\zeta_i)}}.
\end{equation}
This vector potential generates a sinusoidal magnetic field as
\begin{equation}
B_x=B_0\sin(y/L),\ \ \ B_y=-B_0\sin(x/L).
\end{equation}
Equations (\ref{distribution}) and (\ref{vector}) provide a self-consistent configuration of relativistic plasmas. 
It is found that this equilibrium is a relativistic extension of the equilibrium proposed in the previous paper \cite{SS08}.  

\begin{figure}
\includegraphics[bb= 0 0 200 200]{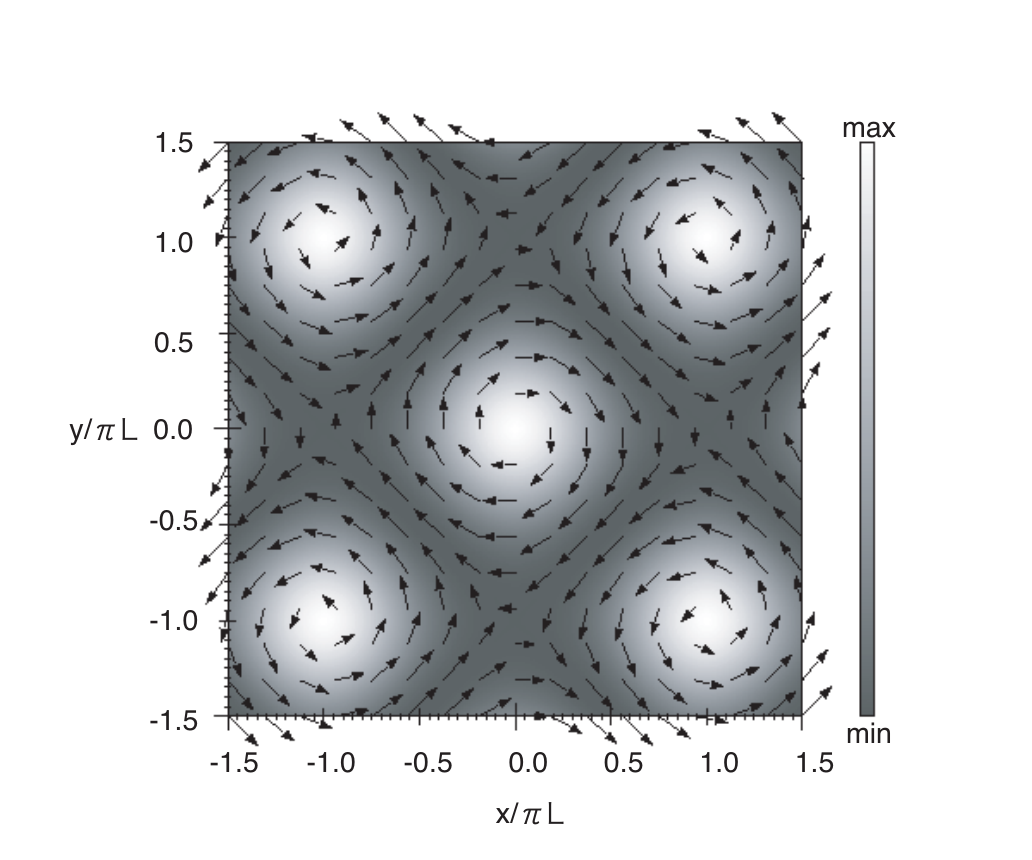}
\caption{The configuration of the two-dimensional relativistic Vlasov-Maxwell equilibrium derived by applying the method in this paper. 
The gray scale represents the density distribution of electrons (or ions) in arbitrary units and the arrows represent the magnetic fields.}
\label{figure1}
\end{figure}

Figure \ref{figure1} illustrates the thus derived configuration of equilibrium.
The gray scale and the arrows represent the density distribution $g_{j,0}^\mathrm{even}$ of electrons or ions in arbitrary units and the magnetic field, respectively. 
As in the non-relativistic counterpart \cite{SS08}, we can see that the current filaments lie along the $z$-axis and generate the magnetic fields around themselves. 
Each filament is surrounded by four filaments that carry anti-parallel currents. 

\begin{figure}
\begin{center}
\includegraphics[bb= 0 0 200 200]{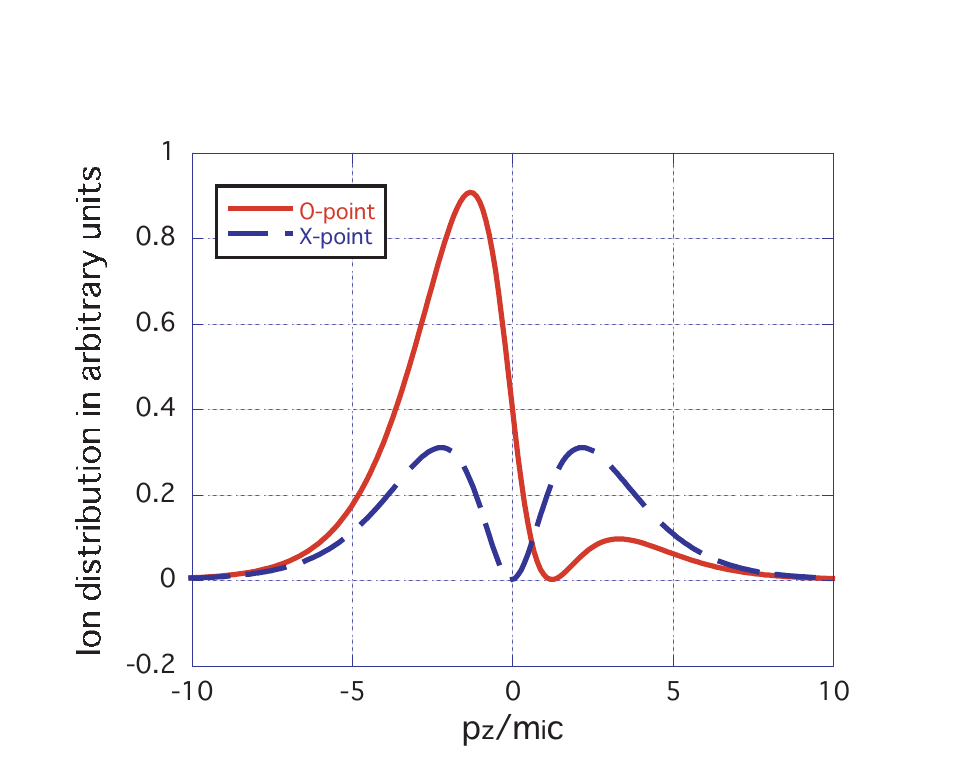}
\caption{(Color online) The momentum distributions of ions at the O-point (solid line) and the X-point (dashed line).}
\label{figure2}
\end{center}
\end{figure}

Here, I focus on the momentum distributions at two points. 
One is the center of a filament (referred to as the O-point) and the other is a middle point between the filaments carrying parallel currents (referred to as the X-point). 
The magnetic field at the X-point is sheared like the Harris sheet equilibrium. 
Figure \ref{figure2} shows the momentum distributions of ions at the X and O-points. 
The values of parameters are as follows; 
$B_0=0.1m_e\omega_e/e$, where $\omega_e$ is the electron plasma frequency, and $kT_i=kT_e=m_ic^2$. 
From these values, one obtains
\begin{equation}
\frac{K_3(\zeta_i)}{\zeta_iK_2(\zeta_i)}\simeq 0.0005,\ \ \ 
\frac{m_e^2K_3(\zeta_e)}{m_i^2\zeta_eK_2(\zeta_e)}\simeq 0.
\end{equation}
So I have assumed $C=0.0005$. 
While a symmetric double-peak distribution appears at the X-point due to $A_z=0$, an asymmetric one is achieved at the O-point. 

In contrast to ions, the momentum distribution of electrons has a symmetric double-peak both at the X and O-points, because the term $e^2A_z^2/m_i^2c^4(<0.04m_e^2\omega_e^2L^2/m_i^2c^4)$ is much smaller than $C(=0.0005)$.

Finally, I make a remark on one of the relativistic effects. 
In comparison with the non-relativistic counterpart, the distribution tends to have a relatively shallow slope in the high-energy regime, which comes from the difference in the equilibrium distribution. 
While the Maxwell-Boltzmann distribution is proportional to $\exp(-p_z^2/2m_jk_\mathrm{B}T)$, the Maxwell-J\"uttner distribution is proportional to $\exp (-cp_z/k_\mathrm{B}T)$ in the limit of $p_z/m_jc \gg 1$. 
In other words, the relativistic one has more high-energy particles than the non-relativistic does. 

\section{\label{conc}CONCLUSIONS}
In this paper, I have developed a novel method to construct stationary solutions of the relativistic Vlasov-Maxwell system. 
By applying the method, a new two-dimensional equilibrium, which is a relativistic extension of the previous work \cite{SS08}, is proposed. 
A comparison of the non-relativistic and the relativistic equilibrium is done. 
As well as the non-relativistic equilibrium, a sheared magnetic field is generated in the relativistic one. 
On the other hand, they show different behaviors in the high-energy regime. 
It may provide an initial setup for investigations of the relativistic collisionless magnetic reconnection that takes into account three-dimensional effects. 
It will be intriguing to investigate the stability of the equilibrium derived in this paper. 

\begin{acknowledgments}
I am grateful to Toshikazu Shigeyama for his useful and constructive suggestion on the manuscript. 
This work is supported in part by Grant-in-Aid for Scientific Research (16540213) from the Ministry of Education, Culture, Sports, Science, and Technology of Japan and JSPS (Japan Society for Promotion of Science) Core-to-Core Program ``International Research Network for
Dark Energy''.
\end{acknowledgments}

\appendix
\section{Properties of the Modified Bessel Functions of Second Kind}
In the next section, I often use the properties of the modified Bessel functions of second kind. 
Therefore, I review the properties in this section \cite{bessel}. 

The $n$th-order modified Bessel functions of second kind $K_n(x)$ is defined by the following integral;
\begin{equation}
K_n(x)=\int ^\infty_0 e^{-x\cosh\theta}\cosh(n\theta)d\theta.
\end{equation}
It is known that $K_n(x)$ and $K_{n+1}$ satisfy the following recurrence formula; 
\begin{equation}
\frac{d}{dx}\left(\frac{K_n}{x^n}\right)=-\frac{K_{n+1}}{x^n}.
\label{dK}
\end{equation}
After some algebraic manipulations, $K_n(x)$ is expressed in another form as
\begin{equation}
K_n(x)=\left(\frac{x}{2}\right)^n\frac{\Gamma(1/2)}{\Gamma(n+1/2)}
\int^{\infty}_0e^{-x\cosh t}\sinh^{2n}tdt.
\label{Kn}
\end{equation}
Dividing both sides of this equation by $x^n$ and differentiating with respect to $x$ yields the relation
\begin{equation}
K_{n+1}(x)=\left(\frac{x}{2}\right)^n\frac{\Gamma(1/2)}{\Gamma(n+1/2)}
\int^\infty_0e^{-x\cosh t}\cosh t\sinh^{2n}tdt.
\label{Kn+1}
\end{equation}
In the limit of $x\gg 1$, $K_n(x)$ reduces to
\begin{equation}
K_n(x)\simeq \sqrt{\frac{\pi}{2x}}e^{-x}.
\label{Ksim}
\end{equation}

\section{\label{Orth}Construction fo Orthogonal Polynomials}
In this section, I construct two special orthogonal polynomial series used in this paper. 
\subsection{Some useful integrals}
In constructing the polynomials, two integrals are used again and again. These integrals are introduced beforehand. 

The first one is
\begin{equation}
I_{n}^\mathrm{odd}=\int^\infty_{-\infty}\int^\infty_{-\infty}\int^\infty_{-\infty}
\hat{p}_z^{2n}\exp\left[-\zeta\sqrt{1+\hat{p}_x^2+\hat{p}_y^2+\hat{p}_z^2}\right]
\frac{d\hat{p}_xd\hat{p}_y\hat{p}_z}
{\sqrt{1+\hat{p}_x^2+\hat{p}_y^2+\hat{p}_z^2}}.
\label{integral_odd}
\end{equation}
where $n$ is an integer and $\zeta$ is an independent variable. 
Introducing the spherical coordinates $(p,\theta,\phi)$ as
\begin{equation}
\hat{p}_x=p\sin\theta\cos\phi,\ \ \ 
\hat{p}_y=p\sin\theta\sin\phi,\ \ \ 
\hat{p}_z=p\cos\theta,
\label{sph}
\end{equation}
the integrations with respect to $\theta$ and $\phi$ can be performed to obtain
\begin{equation}
I_{n}^\mathrm{odd}=\frac{4\pi}{2n+1}\int^\infty_{0}
p^{2n+2}\exp\left[-\zeta\sqrt{1+p^2}\right]
\frac{dp}{\sqrt{1+p^2}}.
\end{equation}
Furthermore, a change of the integral variable from $p$ to $t$ defined as
\begin{equation}
p=\sinh t,
\label{new_var}
\end{equation}
leads to
\begin{equation}
I_{n}^\mathrm{odd}=\frac{4\pi}{2n+1}\int^\infty_{0}
e^{-\zeta\cosh t}\sinh^{2n+2}tdt.
\end{equation}
Using the relation (\ref{Kn}), the integral is expressed in terms of $K_{n+1}$ as
\begin{equation}
I_{n}^\mathrm{odd}=\frac{2^{n+2}\pi\Gamma(n+1/2)}{\zeta^{n+1}\Gamma(1/2)}
K_{n+1}(\zeta)
\label{Iodd}
\end{equation}

The second integral is
\begin{equation}
I_{n}^\mathrm{even}=\int^\infty_{-\infty}\int^\infty_{-\infty}\int^\infty_{-\infty}
\hat{p}_z^{2n}
\exp\left[-\zeta\sqrt{1+\hat{p}_x^2+\hat{p}_y^2+\hat{p}_z^2}\right]
d\hat{p}_xd\hat{p}_yd\hat{p}_z.
\label{integral_even}
\end{equation}
Again, from the relations (\ref{sph}) and (\ref{new_var}), this integral is converted to 
\begin{equation}
I_{n}^\mathrm{even}=\frac{4\pi}{2n+1}\int^\infty_{0}e^{-\zeta\cosh t}
\sinh^{2n+2} t\cosh tdt
\end{equation}
Using the relation (\ref{Kn+1}), the integral is expressed in terms of $K_{n+2}$ as
\begin{equation}
I_{n}^\mathrm{even}=\frac{2^{n+2}\pi\Gamma(n+1/2)}{\zeta^{n+1}\Gamma(1/2)}
K_{n+2}(\zeta)
\label{Ieven}
\end{equation}

\subsection{Even orthogonal polynomials}
I construct the polynomial series $S^\mathrm{even}_n(\hat{p}_Z)$. 
This polynomial must satisfy two conditions; 
(1) These are even functions. 
(2) The set $S^\mathrm{even}_n$ forms an orthogonal basis in the even function space, in other words, each polynomials satisfy the following orthogonality relation,
\begin{equation}
\int^\infty_{-\infty}\int^\infty_{-\infty}\int^\infty_{-\infty}
S^\mathrm{even}_m(\hat{p_z})S^\mathrm{even}_n(\hat{p_z})
\exp\left[-\zeta\sqrt{1+\hat{p}_x^2+\hat{p}_y^2+\hat{p}_z^2}\right]
d\hat{p}_xd\hat{p}_yd\hat{p}_z
\propto \delta_{mn},
\label{orth_even}
\end{equation}
where I choose $\exp\left[-\zeta\sqrt{1+\hat{p}_x^2+\hat{p}_y^2+\hat{p}_z^2}\right]$ as the weight function and $\delta_{mn}$ represents Kronecker's delta. 
Then we define the polynomials $S^\mathrm{even}_n$ as
\begin{equation}
S^\mathrm{even}_0=1,\ \ \ 
S^\mathrm{even}_n=\hat{p}_z^{2n}+\sum_{k=0}^{n-1}A_{nk}^\mathrm{even}S^\mathrm{even}_k,
\label{deff_even}
\end{equation}
where $A_{nk}^\mathrm{even}$ are coefficients determined by the orthogonality relation (\ref{orth_even}). 
Multiplying $S^\mathrm{even}_n$ in (\ref{deff_even}) by 
$S^\mathrm{even}_m\exp\left[-\zeta\sqrt{1+\hat{p}_x^2+\hat{p}_y^2+\hat{p}_z^2}\right] (m<n)$ and integrating it with respect to $\hat{p}_x$, $\hat{p}_y$, and $\hat{p}_z$, the following expression for $A^\mathrm{even}_{nk}$ is obtained;
\begin{equation}
A_{nm}^\mathrm{even}=-\frac{\int^\infty_{-\infty}\int^\infty_{-\infty}\int^\infty_{-\infty}\hat{p}_z^{2n}S^\mathrm{even}_m\exp\left[-\zeta\sqrt{1+\hat{p}_x^2+\hat{p}_y^2+\hat{p}_z^2}\right]d\hat{p}_xd\hat{p}_yd\hat{p}_z}
{\int^\infty_{-\infty}\int^\infty_{-\infty}\int^\infty_{-\infty}(S^\mathrm{even}_m)^2
\exp\left[-\zeta\sqrt{1+\hat{p}_x^2+\hat{p}_y^2+\hat{p}_z^2}\right]
d\hat{p}_xd\hat{p}_yd\hat{p}_z}.
\end{equation}
Using this equation, one can determine the $n$th order polynomial $S^\mathrm{even}_n$ by induction. 
For example, the denominator and numerator of $A_{10}^\mathrm{even}$ become $I^\mathrm{even}_0$ and $I^\mathrm{even}_1$, respectively. 
Thus $A_{10}^\mathrm{even}(=S^\mathrm{even}-\hat{p}_z^2)$ is expressed as
\begin{equation}
A_{10}^\mathrm{even}=-\frac{K_3(\zeta)}{\zeta K_2(\zeta)}.
\end{equation}
In such a way, the even polynomials $S^\mathrm{even}_n$ are obtained from the lower order polynomials $S^\mathrm{even}_m (m<n)$. 
The expression of the first three are
\begin{subequations}
\begin{eqnarray}
S^\mathrm{even}_0&=&1,\\
S^\mathrm{even}_1&=&\hat{p}_z^2-
\frac{K_3(\zeta)}{\zeta K_2(\zeta)},\\
S^\mathrm{even}_2&=&\hat{p}_z^4-\frac{3}{\zeta}\frac{5K_5(\zeta)K_2(\zeta)-K_4(\zeta)K_3(\zeta)}
{3K_4(\zeta)K_2(\zeta)-[K_3(\zeta)]^2}
\left[\hat{p}_z^2-\frac{K_3(\zeta)}{\zeta K_2(\zeta)}\right]
-\frac{3K_4(\zeta)}{\zeta^2K_2(\zeta)}.
\end{eqnarray}
\label{even123}
\end{subequations}
\subsection{Odd orthogonal polynomials}
The odd polynomial series $S^\mathrm{odd}_n(\hat{p}_z)$ are supposed to  satisfy the following orthogonal relation,
\begin{equation}
\int^\infty_{-\infty}\int^\infty_{-\infty}\int^\infty_{-\infty}
S^\mathrm{odd}_m(\hat{p_z})S^\mathrm{odd}_n(\hat{p_z})
\exp\left[-\zeta\sqrt{1+\hat{p}_x^2+\hat{p}_y^2+\hat{p}_z^2}\right]
\frac{d\hat{p}_xd\hat{p}_yd\hat{p}_z}
{\sqrt{1+\hat{p}_x^2+\hat{p}_y^2+\hat{p}_z^2}}
\propto \delta_{mn}.
\label{orth_odd}
\end{equation}
Note that the weight function differs from that of the even polynomial series. 
Then I define the polynomials $S^\mathrm{odd}_n$ as
\begin{equation}
S^\mathrm{odd}_0=\hat{p}_z,\ \ \ 
S^\mathrm{odd}_n=\hat{p}_z^{2n+1}+\sum_{k=0}^{n-1}A_{nk}^\mathrm{odd}S^\mathrm{odd}_k,
\end{equation}
where $A^\mathrm{odd}_{nk}$ are coefficients determined by the same procedure as the even polynomial series. 
Thus, $A^\mathrm{odd}_{nm}$ are expressed as
\begin{equation}
A^\mathrm{odd}_{nm}=
-\frac{\int^\infty_{-\infty}\int^\infty_{-\infty}\int^\infty_{-\infty}\hat{p}_z^{2n+1}S^\mathrm{odd}_m\exp\left[-\zeta\sqrt{1+\hat{p}_x^2+\hat{p}_y^2+\hat{p}_z^2}\right]d\hat{p}_xd\hat{p}_yd\hat{p}_z}
{\int^\infty_{-\infty}\int^\infty_{-\infty}\int^\infty_{-\infty}(S^\mathrm{odd}_m)^2
\exp\left[-\zeta\sqrt{1+\hat{p}_x^2+\hat{p}_y^2+\hat{p}_z^2}\right]
d\hat{p}_xd\hat{p}_yd\hat{p}_z}.
\end{equation}
From this equation and the relation (\ref{Iodd}), one can derive the expression of the odd polynomial series. 
The first three are
\begin{subequations}
\begin{eqnarray}
S^\mathrm{odd}_0&=&\hat{p}_z,\\
S^\mathrm{odd}_1&=&\hat{p}_z^3-\frac{3K_3(\zeta)}{\zeta K_2(\zeta)}\hat{p}_z,\\
S^\mathrm{odd}_2&=&\hat{p}_z^5-\frac{5}{\zeta}
\frac{7K_5(\zeta)K_2(\zeta)-3K_4(\zeta)K_3(\zeta)}
{5K_4(\zeta)K_2(\zeta)-3[K_3(\zeta)]^2}
\left[\hat{p}_z^3-\frac{3K_3(\zeta)}{\zeta K_2(\zeta)}\hat{p}_z\right]
-\frac{15K_4(\zeta)}{\zeta^2K_2(\zeta)}\hat{p}_z.
\end{eqnarray}
\label{odd123}
\end{subequations}
\subsection{Properties}
Here I list some properties of the two orthogonal polynomial series derived above. 
Because the products $S^\mathrm{even}_nS^\mathrm{odd}_m$ are odd function with respect to $\hat{p}_z$, the following integrals vanish;
\begin{subequations}
\begin{eqnarray}
\int^\infty_{-\infty}\int^\infty_{-\infty}\int^\infty_{-\infty}
S^\mathrm{even}_nS^\mathrm{odd}_m\exp\left[-\zeta\sqrt{1+\hat{p}_x^2+\hat{p}_y^2+\hat{p}_z^2}\right]d\hat{p}_xd\hat{p}_yd\hat{p}_z&=&0,
\label{vanisha}\\
\int^\infty_{-\infty}\int^\infty_{-\infty}\int^\infty_{-\infty}
S^\mathrm{even}_nS^\mathrm{odd}_m\exp\left[-\zeta\sqrt{1+\hat{p}_x^2+\hat{p}_y^2+\hat{p}_z^2}\right]\frac{d\hat{p}_xd\hat{p}_yd\hat{p}_z}{\sqrt{1+\hat{p}_x^2+\hat{p}_y^2+\hat{p}_z^2}}&=&0.
\label{vanishb}
\end{eqnarray}
\label{vanish}
\end{subequations}

Next, the derivatives of the polynomials can be expanded by $S^\mathrm{odd}_n$; 
\begin{equation}
\frac{dS^\mathrm{even}_n}{d\hat{p}_z}=
\sum_{k=0}^{n-1}B^\mathrm{even}_{nk}S^\mathrm{odd}_k.
\label{dSeven}
\end{equation}
Because the derivatives of $S^\mathrm{even}_n$ is an odd function. 
On the other hand, the derivatives of $S^\mathrm{odd}_n$ can be expanded by $S^\mathrm{even}_n$;
\begin{equation}
\frac{dS^\mathrm{odd}_n}{d\hat{p}_z}=
\sum_{k=0}^{n}B^\mathrm{odd}_{nk}S^\mathrm{even}_k.
\label{dSodd}
\end{equation}
The expressions of $B^\mathrm{even}_{nk}$ and $B^\mathrm{odd}_{nk}$ are obtained by directly differentiating the expressions (\ref{even123}), (\ref{odd123}), and their extensions for larger $n$.

\newpage

\end{document}